\documentstyle[12pt]{article}

\newcommand{\req}[1]{(\ref{#1})}

\begin{document}

\begin{titlepage}

\vspace{.5in}
\begin{flushright}
22th November 1998 \\
gr-qc/9811076 \\
\end{flushright}
\vspace{.5in}

\begin{center}
{\large\bf Boundary Terms and Noether Current \\
 of Spherical Black Holes}
\\

\bigskip
\bigskip

M.  C.  Ashworth
\footnote{\it email: ashworth@th.phys.titech.ac.jp}\\
{\small\it Department of Physics}\\
{\small\it Tokyo Institute of Technology}\\
{\small\it Ohokayama, Meguro, 152-8551, JAPAN}\\

\bigskip

Sean A. Hayward
\footnote{\it email: hayward@yukawa.kyoto-u.ac.jp} \\
{\small\it Yukawa Institute for Theoretical Physics}\\
{\small\it Kyoto University} \\
{\small\it Kitashirakawa, Sakyo-ku, Kyoto 606-8502, JAPAN}\\

\end{center}

\vspace{.5in}
\begin{center}
{\bf Abstract}\\
\end{center}

\begin{center}

\begin{minipage}{5in}
{\small
We consider two proposals for defining black hole entropy in
spherical symmetry, where the horizon is defined locally as
a trapping horizon.  The first case, boundary terms in a
dual-null form of the reduced action in two dimensions,
gives a result that is proportional to the area.
The second case, Wald's Noether current method, is generalized to dynamic
black holes, giving an entropy that is just the area of the trapping
horizon.
These results are compared with a generalized first law of
thermodynamics.

}
\end{minipage}
\end{center}
\end{titlepage}
\addtocounter{footnote}{-2}

\section{Introduction}

There has been much interest in the problem of black hole entropy,
ever since the famous argument of Bekenstein was introduced
to save the second
law of thermodynamics: in order for entropy not be lost
into a black hole, the black hole must have entropy.
Then, Hawking \cite{HK} discovered that black holes radiate with
a black body spectrum at a temperature of $T = \kappa / 2 \pi$
where $\kappa$ is the surface gravity at the horizon.  These two
discoveries led
to the study of black hole thermodynamics.  In particular,
stationary black holes were found \cite{Carter}
to have a first law or energy balance equation
$\delta m = \kappa  \delta A / 8 \pi$ plus work terms.
To date, most of the work  in this field has been done for
stationary or (in some sense) quasi-stationary black holes.  This
should properly be thought of as thermostatics and not
thermodynamics. The dynamical case contains many technological and
definitional problems which are recently starting to be resolved.  For
example, a generalized first law of thermodynamics
has been suggested in \cite{Sean3}.

However, black hole thermodynamics and thermostatics
have a footing only in
the classical regime. There is as yet no clear statistical mechanical
origin for entropy of a black hole. Do the states that
make up this entropy lie inside the black hole, throughout the
space-time, on the event horizon, or on a local horizon? What quantum states
compose this entropy?  Do these states preserve the
``information'' collected in the black holes by
given correlation functions with
the radiated energy?  Is there a remnant quanta at the
end of evaporation? As such, black hole entropy is one of the
leading testing grounds for quantum gravity.  One of the key tests
to a  successful quantum
theory is the ability to predict the area law, which to date
has been tested in various cases in the classical limit.
There have been many suggestions as to where this entropy might come
from and ways of calculating it: Noether current \cite{Wald1},
D-branes \cite{dbrane}, spin systems \cite{spin}, entanglements
\cite{entangle},
and surface terms that break the symmetry of the system
\cite{surface} \cite{surface2}.
Each of these areas shows promising results but many of them contain
limitations.  For example, the Noether current needs a bifurcation
surface in the system to define the entropy \cite{Wald1}.
In the case of
boundary terms, the only quantum calculations done to date have been in
2+1 dimensions \cite{Carlip1} and also
rely heavily on string techniques that are difficult
generalize to higher dimensions. D-branes require
higher dimensionality of space-time.

We wish to look into the problem of black hole entropy
in the special case of spherical symmetry.
There are many advantages to
this case. Firstly, spherical symmetry
simplifies many technical problems but still allows enough degrees
of freedom to be interesting and dynamical.
Stationary black holes are fairly
well understood.  However, generally black holes will
go through some dynamical period in which the
horizon area increases or decreases as matter is accumulated into
the black hole or as energy is radiated away,
for instance by the Hawking process.
It is this dynamical setting that we are most
interested in studying. Secondly, in spherical symmetry, there is
preferred time direction given by the Kodama vector
\cite{Kodama2}.  In the stationary case, the metric admits a
time-like Killing vector from which we define
time (an important step in quantizing a system).
The Kodama vector is a natural analogue of a Killing vector \cite{Sean3}
which can be used in a dynamical setting.
Related to this choice of time, we can define a
local energy where normally only a global energy can be defined.
Thirdly, there is a dynamical local definition of the outer
surface of a black hole. In stationary systems, the
horizon is a Killing horizon, which coincides with both the event horizon
and the apparent horizon.
In a general case, the locally defined
apparent horizon is not the same as the event horizon. So there arises a
problem of which horizon gives the black hole entropy.
In the case of
spherical symmetry it is particularly simple to define a local horizon as a
trapping horizon \cite{Sean1}, a hypersurface foliated by marginal surfaces.
This definition does not depend on
any global characteristics such as the asymptotic behavior of the
space-time and is a natural candidate for the outer surface of the
back hole.

In this paper, we will look at two methods of defining black hole
entropy for the case of spherical symmetry.
In the second section, a reduced
dimensional action is derived by integrating out the angular
degrees of freedom on the sphere of symmetry, keeping the surface
terms.  It is from these surface terms that entropy is proposed to
arise \cite{Hawking1} \cite{Hawking2}.  In the dual-null form
of the action, we
indeed get an edge term that is proportional to the area, suggesting a
possible dynamical definition of the entropy. In the third
section, we look at the Noether currents of the action. In a
n-form structure, Wald \cite{Wald1} found that the conserved
charge associated with the Killing fields on the black hole when
suitably rescaled gives the entropy of the stationary black hole.
This method is easily generalizable to other actions. So along these
lines, we begin by looking at
the normal conserved charges of the spherically symmetric Einstein
action. From these charges it is possible to identify the energy
and write down an energy balance or first law from which an entropy
can be defined \cite{Sean3}.  Then, we generalize Wald's Noether
charge to the trapping horizon for both the reduced action and the
full four-dimensional action.  The resulting dynamical conserved
charge takes a similar form to the stationary case and reduces exactly
to Wald's charge in the stationary limit.  This charge when rescaled
in the same way as the Wald charge gives an entropy that is one-quarter of
the area of the trapping horizon in the case of Einstein gravity.
In the derivation, it is not
required to have a bifurcation surface in any sense.
Then, in this dynamical setting,
the question of what this conserved charge might be is
addressed.

\section{Reduced Action with Boundary Terms}

In deriving the equation of motion from an action,
a boundary term must sometimes
be included to cancel a total derivative when integrating by
parts.  It has been proposed \cite{Hawking1} that this
surface term gives rise to
the entropy of the black hole.  It indeed gives the area law in
the case of the Euclidean black hole \cite{Banados1}.
It is these surface terms which we wish to study in the spherically
symmetric case where the boundary of the black hole can be defined
dynamically as the trapping horizon.  The resulting edge term in
the reduced action defined below is proportional to the area.  As such,
it is a possible candidate for a dynamical definition of the
entropy of a black hole.

In the case of
spherical symmetry, it possible to write an effective two-dimensional
action by integrating over the angular directions of the sphere.
Let us look at this action and
the resulting boundary term. For this derivation,
we start with the action with boundary terms in four-dimensional
space-time,

\begin{equation}
S =\int_{M}  \sqrt{g} R^{(4)} d^4 x
+ 2 \int_{\partial M} \sqrt{g} K^{(4)} d^3 x.
\label{action}
\end{equation}

\noindent
In the case of sphere symmetry, we can write the line element as

\begin{equation}
ds^2  = r^2 d^2 \Omega + g_{ij} d x^i d x^j
\end{equation}

\noindent
where $g^{ij}$ is the induced metric on the remaining two
dimensional sub-manifold.  The manifold is then foliated by
spheres labelled by the coordinates $\{ x_i \}$ which have an area
$A$. The areal radius is then defined by $r=\sqrt{A/ 4 \pi}$.
The scalar curvature in four dimensions can
be written in terms of the curvature of the two-dimensional normal
sub-manifold
and a term involving $r$,

\begin{equation}
R^{(4)} = R^{(2)} + { \left( 2 - 2  g^{ij} \nabla_i \nabla_j (r^2)
+ 2 g^{ij} \nabla_i r \nabla_j r \right) \over r^2}.
\end{equation}

\noindent
The boundary surface will also be chosen to respect the
spherical symmetry, giving rise to a natural definition of the
boundary on the sub-manifold.
The normal to the surface in coordinate
representation takes the
form $n^a = \{ 0,0,n^i(x) \}$, where the first two coordinates are
the angular directions, and $n^i(x)$ is independent of the angular
coordinates. With this definition,
the extrinsic curvature can also be written in terms of
its two-dimensional counterpart and the areal radius,

\begin{equation}
K^{(4)} = K^{(2)} + { n^j \nabla_j (r^2) \over r^2 }.
\end{equation}

\noindent
The terms in the above action \req{action} can be replaced by
their two-dimensional counterparts suggesting that an effective or
reduced dimensional action can be written
on the two-dimensional normal manifold.  We derive this action
by integrating out the angular dependence, in essence averaging
over them. Substituting in the above express for the
curvature and integrating the angular directions,
we get a reduced action in two dimensions,

\begin{eqnarray}
S & =& 2 \pi \int_{M} \sqrt{g} \left( r^2  R^{(2)}
+ 2  - 2  g^{ij} \nabla_i \nabla_j (r^2)
+ 2 g^{ij} \nabla_i r \nabla_j r \right) d ^2 x
\nonumber \\[1ex]
& & + 4 \pi  \int_{\partial M} \sqrt{g} \ r^2 K^{(2)}
+  n^j \nabla_j (r^2) d^1 x.
\end{eqnarray}

\noindent
Note: the third term in the reduced action is just a
total derivative and does not affect the dynamics.
So removing this term as a boundary term, it cancels the
second term in the surface action, leaving us with

\begin{equation}
S  = 2 \pi \int_{M} \sqrt{g} \left( r^2  R^{(2)}
+ 2 + 2 g^{ij} \nabla_i r \nabla_j r \right) d ^2 x
+ 4 \pi  \int_{\partial M} \sqrt{g} \ r^2 K^{(2)} d^1 x.
\label{reduce}
\end{equation}

\noindent
This action takes a similar form as
the string motivated dilaton action \cite{Dilaton} where the areal
radius replaces the dilaton ($r^2 = e^{-\phi}$).
However, it contains a two-dimensional cosmological constant like
term (the second term in \req{reduce}).  This term breaks the
conformal symmetry, which is to be expected since the original
four-dimensional action was not conformally invariant.

For the reduced action \req{reduce}, we get two sets of
equations of motion, one from
varying the two-dimensional metric and the other from varying $r$.
As we will see, these equations of motion replicate the
equations of motion coming for the original four-dimensional
action \req{action}.
In the derivation\footnote{See Wald \cite{Wald} appendix E }
of the four-dimensional surface
term $\int K$, the
metric on the boundary is taken to be fixed.  This means that the
induced metric and $r$, which replaces $g_{\theta \theta}$
and $g_{\phi \phi}$ should also be fixed on the boundary,
thus to be consistent we should also have
Dirichlet boundary conditions for $r$. The reduced
action with respect to the variation in the metric gives,
\begin{eqnarray}
\delta S & = & 2 \pi \int_{M} \sqrt{g} \delta g^{ij} \left[
-{1 \over 2} g_{ij}( 2 - 2 g^{kl} \nabla_k r \nabla_l r +
4 g^{kl} r \nabla_k \nabla_l r )
-2 r \nabla_i \nabla_j r \right]
\nonumber \\[1ex]
& & + 2 \pi \int_{\partial M} \sqrt{h}
\left[ (g^{ik} g^{jl} - g^{ij} g^{kl})
(n_i r^2 \nabla _j ( \delta g_{kl}) - n_j \nabla _i (r^2) \delta
g_{kl})\right]
\nonumber \\[1ex]
&& + 4 \pi \int_{\partial M} \sqrt{h}
r^2 n^i h^{jk} \nabla_i (\delta g_{jk}),
\label{reduce2}
\end{eqnarray}
where we have made use of the fact that $R_{ab} = 1/2 g_{ab} R$
in two dimensions. The metric $h_{ab}$ is the induced metric on the
one dimensional surface.  For the boundary conditions
$\delta g_{ab} =0 $ on the surface, the surface terms cancel.
For the case that only $\delta h_{ab}=0$, there exists
a gauge transformation on the surface that will make
$\delta g_{ab}=0$, thus fixing part of the gauge on the surface.
The volume term then gives the equations of motion,
\begin{equation}
{1 \over 2} g_{ij}( 2 - 2 g^{kl} \nabla_k r \nabla_l r +
4 r g^{kl} \nabla_k \nabla_l r )
-2 r \nabla_i \nabla_j r = 8 \pi T^{(2)} _{\ ij}
\label{eqom1}
\end{equation}
where $T^{(2)} _{\ ij}$ is the reduced stress-energy tensor of
the matter fields.

Now let us consider the variation of the reduced action
\req{reduce} with respect to the areal radius
$r$.
\begin{equation}
\delta S  =  2 \pi \int_{M} \sqrt{g}
\ \delta r \left[2 R - 4 g^{ij} \nabla_i \nabla_j r \right]
+ 4 \pi \int_{\partial M} \sqrt{h}
\ \delta r \left[ 2K + 2 n^i \nabla_i r  \right].
\end{equation}
For the boundary condition $\delta r =0$, the surface term
drops and we are left with the equation of motion from
the volume term,
\begin{equation}
2r R - 4 g^{ij} \nabla _i  \nabla_j r = - 4 \pi \rho
\label{eqom2}
\end{equation}
where $\rho = {1 \over 4 \pi} \delta S_{\rm matter} / \delta r$
which is related to the angular stress-energy terms from the
original four-dimensional theory, as we will see below. To try to
generalize to other boundary conditions is difficult because of
the non-linear way in which the components of the metric tensor
interact.
To complete the equations of
motion, we need to consider the form of the matter action in the
reduced two-dimensional theory.

Of course it is possible to consider complicated and more
general matter actions, however, to
simplify matters, let us only consider diffeomorphism invariant
actions that have the form ${\cal L} (\phi, \nabla \phi, g)$ where
the $\phi$ is the set of matter fields.  In this paper,
we will not consider
matter coupled to the curvature.  Let us
start with the Klein-Gordan action,
\begin{equation}
S_m = \int \sqrt{g} ( g^{ab} \partial _a \phi \partial _b \phi +
m^2 \phi ^2 ) d^4 x.
\end{equation}
For notational purposes, tensor and operators labelled by the
beginning of the alphabet $(a,b,c,\ldots)$ will be four-dimensional,
while their counterparts in two dimensions will be labelled by
the middle of the alphabet $(i,j,k,\ldots)$.
If the fields are not dependent on the angular coordinates,
then the action is just,
\begin{equation}
S_m = 2 \pi \int \sqrt{g^{(2)}} r^2 ( g^{ij} \partial _i \phi
\partial _j \phi + m^2 \phi ^2 ) d^2 x.
\end{equation}
Generically, the action will take the form
\begin{equation}
S_m = 2 \pi \int \sqrt{g} r^2 {\cal L}'
\end{equation}
With spherical symmetry,
the four-dimensional stress-energy tensor takes block
diagonal form, $T_{\theta \theta} d \theta^2
+ T_{\phi \phi} d \phi^2 + T^{(4)}_{\ ij} d x^i d x^j$.
The stress-energy tensor of the reduced action is related to the
restriction of the four-dimensional stress energy by a scaling
factor which is just the area,
\begin{equation}
T^{(2)}_{ij} = r^2 \left[ \partial_i \phi \partial_j \phi -
{1 \over 2} g_{ij}( g^{kl} \partial_k \phi \partial_l \phi
+ m^2 \phi ^2 ) \right]  = 4 \pi r^2 T^{(4)}_{\ ij}
\end{equation}
To recover all the four-dimensional equations of motion, we
must also consider variations with respect to $r$.
This variation of the matter actions given us the
definition of $\rho$ \req{eqom2},
\begin{equation}
\rho = 2 r (g^{ij} \partial_i \phi \partial_j \phi
+ m^2 \phi ^2 ) = - 4 r T_\theta ^\theta
\end{equation}
In the above equations \req{eqom1} and \req{eqom2},
we can see how the original stress-energy tensor enters into the
reduced two-dimensional equation of motion.  In
this form it is easy to see that for a generic action, it will also
take the same form.  The matter action is encoded into
the two-dimensional stress-energy and $\rho$, which is the
angular part of the stress-energy tensor.

The equations of motion are equivalent under diffeomorphisms,
so we are free to choose any set of coordinates.
The above equations of motion \req{eqom1}
and \req{eqom2}, greatly simplify in the case of double-null
coordinates given by
\begin{equation}
g_{ij} = \left( \matrix{
0 &-e^{-f} \cr
-e^{-f} & 0 } \right) .
\end{equation}
In this gauge choice, we get the same
equations of motion as \cite{Sean2},
\begin{eqnarray}
&& \partial_{\pm} \partial_{\pm} r + \partial_{\pm} f
\partial_{\pm} r = - 4 \pi r T^{(4)}_{\ \pm \pm}
\label{eqm1} \\[1ex]
&& r \partial _+ \partial _
- r + \partial_+ r \partial _- r
+ {1 \over 2} e^{-f} = 8 \pi r^2 T^{(4)}_{\ +-}
\\[1ex]
&& r \partial_+ \partial _- f - 2\partial _+ \partial_- r
= 8 \pi r e^{-f}T_\theta^\theta
\label{eqm2}
\end{eqnarray}

The areal radius $r$  is related to the causal structure
of the space-time.  Following \cite{Sean1},
a sphere is said to be untrapped, marginal or trapped as
$\nabla^a r$ is spatial, null, or temporal.  For $\nabla^a r$
future directed, the trapped (or marginal) surface is said to be
future trapped, and likewise for past trapped surfaces.
A hypersurface that is foliated by
marginal surfaces is called a trapping horizon. A future outer trapping
horizon is proposed to be the outer boundary of a black hole
\cite{Sean2}.  This definition is purely locally defined, and unlike
event horizons or apparent horizons, it does not depend on any global
conditions such as asymptotic flatness.

It seems natural to set boundary conditions for a dynamical black hole
on the trapping horizon, which may be thought of as the
inner boundary of the external space-time. Normally one fixes the
outer boundary as past and future infinity ($\cal J^{\pm}$),
with the boundary
conditions being set by some asymptotic behavior. In
particular, it is possible to determine a total energy of the
system, the Bondi energy or (at spacelike infinity)
the ADM energy \cite{ADM}.
Such energies are defined globally.
However in the spherically
symmetric case, the Misner-Sharp \cite{Miser} energy
can be defined as

\begin{equation}
E= {1 \over 2} r ( 1- \nabla_a r \nabla^a r),
\end{equation}

\noindent
which is a purely local statement.  The various properties of
this energy have been investigated in various
limiting cases \cite{Sean2}.  It short,
it represents a local active
gravitational energy. On the trapping horizon, $\nabla^a r$ is
null.  Therefore, on the horizon, $2E = r$,
which generalizes the normal definition of the radius of
the Schwarzschild event horizon to dynamic black holes.

With these local definitions of the boundary of a black hole and
gravitational energy, we would like to further investigate the
entropy of this system.  As stated before,
it is believed that one of the
possible sources of entropy comes from the surface term in the
above action \req{action} \cite{Banados1}.  Following these lines, let us
study the boundary in the reduced action,  using a trapping
horizon as the inner boundary of the space-time (the outer
boundary of the black hole). This hypersurface is typically spacelike
for a dynamic black hole, but it becomes null for a stationary black hole.
For a null hypersurface the normal vector becomes tangent to the
hypersurface,
so that the usual 3+1 definition of extrinsic curvature becomes an
intrinsic function of the hypersurface.  For this reason and others,
it is easier to process
in a double-null formulation.

Generalized Lagrangian and Hamiltonian theories of double-null systems
were developed in Ref. \cite{dn}
and applied to the Einstein system in Ref. \cite{null}.
The basic idea is to have an action
\begin{equation}
S_{DN}=\int L\, dx^+ dx^-
\end{equation}
where the Lagrangian $L$ is a function of some configuration fields $q$
and two corresponding Lie derivatives $L_\pm q$,
which in the current case reduce to the partial derivatives
$\partial_\pm q=\partial q/\partial x^\pm$.
The Lagrangian for vacuum Einstein theory given in \cite{null} evaluates to
\begin{equation}
L=\partial_+A\partial_-f+\partial_-A\partial_+f
-A^{-1}\partial_+A\partial_-A+8\pi e^{-f}
\end{equation}
in spherical symmetry.
One may check that the corresponding Euler-Lagrange equations yield
the $G_{+-}$ and $G^\theta_\theta$ components of the Einstein equation,
with the $G_{\pm\pm}$ components requiring Lagrange multipliers.
This Lagrangian was obtained from the Einstein-Hilbert Lagrangian
\begin{equation}
S_{EH} =\int_{M}  \sqrt{g} R^{(4)} d^4 x
\end{equation}
by removing boundary terms,
which may be recovered using the expression (3) for $R^{(4)}$
and noting $R^{(2)}=-2e^f\partial_+\partial_-f$, yielding
\begin{equation}
S_{EH}=\int\left(4\partial_+\partial_-A-2A\partial_+\partial_-f
-A^{-1}\partial_+A\partial_-A+8\pi e^{-f}\right)dx^+dx^-.
\end{equation}
A comparison allows one to write
\begin{equation}
S_{DN}=S_{EH}+S_++S_-+S_0
\end{equation}
where we have separated three boundary terms:
\begin{equation}
S_\pm=\int A\partial_\pm f dx^\pm
\end{equation}
are obtained by integrating total derivatives in $\partial_\mp$,
but there is also a double total derivative in $\partial_+\partial_-$
which may be integrated twice to
\begin{equation}
S_0=-4A.
\end{equation}
The fact that this double boundary term is basically the area
suggests a connection with entropy. Comparing with the $3+1$
ADM formalism in \cite{Hawking1}, we also
see their entropy is related to an edge term.
The other boundary terms $S_\pm$ are gauge-dependent
and can be set to zero by taking $x^\pm$ to be affine parameters,
$\partial_\pm f=0$,
on the null hypersurfaces $x^\mp=0$.
Consequently a suggestion for defining entropy is
\begin{equation}
-{S_0\over{16}}.
\end{equation}
Note that this makes sense in the dynamical case,
agreeing with the usual expression $A/4$,
and may be generalized to other theories of gravity
in which a similar decomposition of the dual-null action occurs.

Alternatively, since gauge dependence
has been shown \cite{surface2} to lead to
observables on the edge, so called edge states,
it may be interesting to study the gauge dependent surface terms $S_\pm$.
However,
it is unclear if the above double-null formulation is complete for
two reasons. The first is that the gauge transformation of the null
coordinates are only a small subset of the original gauge
transformations. Related to this, the second reason is that
the equations of motions above \req{eqm1}-\req{eqm2}
should all be reproduced in the double null formulation.
Namely, the equation \req{eqm1}
most be included as a Lagrange multiplier term.  This term might
also lead to an edge observables.

\section{Noether Currents}

Next, we would like to consider the Noether currents of the
reduced action and the original four-dimensional action.
In Wald and Iyer's work \cite{Iyer1},
they have defined an
entropy in terms of the Noether current.  We would like to compare
Wald's currents with the known conversed currents of the
spherically symmetric theory and the
generalized entropy discussed in \cite{Sean3}.

For every symmetry, there is an associated
conserved current (Noether theorem). Gravitational theories
are generally defined from a diffeomorphism
invariant action.  These diffeomorphisms are locally generated by
an arbitrary vector field $\xi ^a$.  So for each of these vector
fields there is an  associated $(n-1)$ form Noether current and a
$(n-2)$ form Noether charge, where $n$ is the dimension of the
manifold on which the action is defined.  Wald
defined an entropy for a stationary black in terms of the integral
of the Noether charge associated with the horizon Killing fields
on the bifurcation surface \cite{Wald1}.  In this definition, it
is required that the surface gravity be normalized to unity.

The Noether
current is defined in terms of the symplectic potential $\Theta$
and the n-form action $L$ \cite{Wald1},

\begin{equation}
J= \Theta(\phi, {\cal L }_{\xi} \phi) - \xi \cdot L
\end{equation}

\noindent
where $\xi$ is the generator of the diffeomorphism.  The symplectic
potential form is defined from variation of the action,

\begin{equation}
\delta L = {\cal E} \delta \phi + d \Theta,
\end{equation}

\noindent
where the $\phi$ is the dynamic fields and ${\cal E}$
is the equations of motion. The potential $\Theta$
is only defined up to a total
derivative.\footnote{See \cite{Iyer1} for a discussion of methods
to choose a given form of the symplectic potential.}

Quoting the results for the pure gravitational action \cite{Iyer1},
the Noether current and charge are
\begin{eqnarray}
J_{abc} & = & {1 \over 8 \pi} \epsilon _{dabc} \nabla_e \left(
\nabla^{[e} \xi ^{d]} \right)
\\[1ex]
Q_{ab} & = & - {1 \over 16 \pi} \epsilon_{abcd} \nabla ^c \xi^d
\end{eqnarray}
where the brackets indicate the anti-symmetric sum
with the convention of Wald \cite{Wald}.
For the diffeomorphism generated by the Killing vector $\xi^a$
which generates the Killing horizon,
\begin{equation}
\epsilon_{abcd}\nabla^c\xi^d=-2\kappa\eta_{ab}
\end{equation}
where $\kappa$ is the surface gravity,
$\eta_{ab}$ is the area form of the spatial surfaces
lying in the horizon, and the space-time volume form is
$\epsilon_{abcd}=2\epsilon_{[ab}\eta_{cd]}$.
Thus the integral of the Noether charge over such a bifurcation
surface is
\begin{equation}
\oint Q = {\kappa A \over 8\pi}.
\label{noeth1}
\end{equation}
Then the entropy is defined as $2\pi$ times this charge with the surface
gravity being normalized by $\kappa = 1$, giving the standard
stationary black hole entropy of quarter of the area.

Before looking at this in the spherically symmetric case,
let us look at the known conserved currents of this
system.
There are two conserved currents for the spherical system.  The
first is the Kodama vector.
It is defined as a curl of the areal radius,
$k^a = \epsilon^{ab} \nabla _b r$.
The time-like Killing vector that generates time translations
can be replaced by the Kodama vector in the spherically
symmetric case \cite{Sean3}.  It follows from the above definition
that
\begin{eqnarray}
k^a \nabla_a r = 0
\\[1ex]
k^a k_a = {2 E \over r} - 1
\end{eqnarray}
giving the relation between the local energy and this time like
vector.  It also follows that $k$ is a conserved current,
\begin{equation}
\nabla_a k^a  = 0,
\end{equation}
with conserved charge given by the Gauss theorem,
\begin{equation}
V= - \int_{\Sigma} k^a d\Sigma_a
\end{equation}
where $d\Sigma^a$ is the volume form
times a future directed unit normal vector of the
space-like hypersurface
$\Sigma$.  This charge is the areal volume $V= {4 \over 3} \pi r^3$.

{}From the stress-energy tensor we can define two invariants; the
work density,
\begin{equation}
w  =   -{1\over 2} T^{ij}g_{ij}
\end{equation}
and the energy flux (localized Bondi flux)
\begin{equation}
\psi^a = T^{ab} \nabla_b r + w \nabla^a r.
\end{equation}
The energy flux may be replaced by a energy-momentum density along the
Kodama vector, $j^a = T^{ab} k_b$ or equivalently
\begin{equation}
j^a = \epsilon^{ab} \psi_b + w k^a
\end{equation}
This is also a conserved current,
\begin{equation}
\nabla_a j^a  =  0
\end{equation}
which has a conserved charge equal to the energy,
\begin{equation}
E = - \int_{\Sigma} j^a d\Sigma_a.
\end{equation}

A surface gravity for dynamic black holes can be defined \cite{Sean3}
from the Kodama
vector by noting an analogous definition with the Killing vector
,
\begin{equation}
\xi^b \nabla_{[b} \xi_{a]} = \kappa \xi_a.
\end{equation}
Replacing the Killing vector with the Kodama vector,
we get from the equations of motion
\begin{equation}
k^b \nabla_{[b} k_{a]}
 = (E/r^2 - 4 \pi r w)k_a  = \kappa k_a
\end{equation}
on a trapping horizon.
Alternatively, we may define the surface gravity directly from the
Kodama vector as
\begin{equation}
\kappa = \epsilon^{ab}\nabla_a k_b /2.
\label{surgrav1}
\end{equation}

\noindent
Summing up, we have two kinematical quantities $(r,k^a)$, two
gravitational quantities $(E,\kappa)$ and two matter quantities
$(w,\psi^a)$ or $(w,j^a)$. For these definitions, it is possible
to define an
entropy strictly in a thermodynamical setting \cite{HMA} in the same way
that it was first introduced by Clausius \cite{T}.
{}From the equations of motion, we can write an energy balance
equation or first law
\begin{equation}
\nabla_a E = A \psi_a + w \nabla_a V
\end{equation}
The second term on the right side can be thought of
as a type of work, and the first term as
a energy flux term, analogous to the heat flux term in
standard thermodynamics. Using the above definition of the
surface gravity and the equations of motion,
we can rewrite this energy flux as
\begin{equation}
A\psi_a = {\kappa \nabla_a A \over 8 \pi}
+ r \nabla_a \left( {E \over r} \right)
\end{equation}
On any surface where $E/r$ is constant, this last term disappears.
This is the case on a trapping horizon, where $\nabla^a r$ is null
and $2E = r$.
Identifying the temperature with $\kappa / 2 \pi$, the
entropy associated with the black hole trapping horizon is then
$A/4$.

{}From the conserved currents and the equations of motion, we are
able to identify a first law of thermodynamics which enables us to
derive the expected area law for the entropy.  This definition works
in a dynamical setting.  In order to compare results, we wish to
generalize Wald's results \req{noeth1} to this dynamical setting where
the time-like Killing vector is replaced by the Kodama vector and
the surface is defined as the trapping horizon.

So returning to Wald's define of entropy, let us look at the
Noether currents of spherically symmetric gravity.  We have
two actions to consider, the reduced action and the original four
dimensional action. At this point, we switch to Wald's scaling
of the Lagrangian $l = \epsilon R^{(4)} / 16 \pi$.
Let us start with the reduced action \req{reduce}.  From the
derivation of the equations of motion \req{reduce2},
the symplectic potential can be seen to take the form,
\begin{eqnarray}
\Theta_i & = & {1 \over 8}  \epsilon_{ij}
\left[ (g^{jl} g^{km} - g^{jk} g^{lm})
( r^2 \nabla _l ( \delta g_{km}) -
\nabla _j (r^2) \delta g_{km} )  + 4 \nabla_i r  \delta r \right]
\nonumber \\[1ex]
&& + \ {\rm matter \ terms}.
\end{eqnarray}
The variation be generated by the diffeomorphism given are
by $\delta g_{ij} = \nabla _i \xi _j + \nabla _j \xi _i$ and
$\delta r = \nabla _i r \xi^ i$.  For which the current can be
written,
\begin{eqnarray}
j_i & = & {1 \over 4} \epsilon_{ij} \Big[ \nabla_k \left( r^2 \nabla^{[j}
\xi^{k]} - 2 \xi^{[j} \nabla^{k]}( r^2 ) \right)
\nonumber \\[1ex]
&& - \xi_k \left( g^{jk} (2  - 2 \nabla _l r \nabla^l r
+ 4 r \nabla_l \nabla^l r ) - 4 r \nabla^j \nabla^k r \right)
\Big] + \ {\rm matter \ terms}
\end{eqnarray}
For the case of pure gravity or the matter fields governed by a
Klein-Gordon action, the matter fields can be replaced by the
gravitational fields by use of the equations of motion
\req{eqom1}.  This
exchange cancels many of the terms, resulting in
a conserved current that is equivalent to
the dilaton action derived by Wald in \cite{Iyer1} with $e^\phi$
being replaced by $r^2$,
\begin{equation}
j_i = {1 \over 4} \epsilon_{ij} \nabla_k ( r^2 \nabla^{[j}
\xi^{k]} - 2 \xi^{[j} \nabla^{k]}( r^2 ) ),
\end{equation}
The corresponding conserved charge is then
\begin{equation}
q = - {1 \over 8} \epsilon_{ij} ( r^2 \nabla^{i}
\xi^{j} - 2 \xi^{i} \nabla^{j} ( r^2 ) ).
\end{equation}

If we started from the original four-dimensional action, we get
a slightly different form for this conserved current and charge,
\begin{equation}
J_{iab} = {1 \over 8 \pi}
\eta_{ab} \epsilon_{ij} ( r^2 \nabla_k \nabla^{[j}
\xi^{k]} - 2 r  \nabla_k \xi^{[j} \nabla^{k]} r  ),
\end{equation}
\begin{equation}
Q_{ab} = -{1 \over 16 \pi}
\eta_{ab} \epsilon_{ij} r^2 \nabla^i \xi ^j
\end{equation}
where $\eta_{ab}$ is now the area form of the unit sphere
and $\xi$ is restricted to
the tangent space of the reduced two-dimensional
space. We can compare these terms by noting that
\begin{equation}
J_{iab} \cong {1 \over 2 \pi}
\eta_{ab} j_i \qquad \qquad Q_{ab} \cong {1 \over 2 \pi} \eta_{ab} q
\end{equation}
should be equivalent up to total derivatives.
The difference is due to the fact that we have
removed a surface term $\nabla^a \nabla_a r^2$ in the
reduced action.  However, the current and charge are only defined
up to a total derivative anyway \cite{Wald1}.  This once again
brings up an issue that Iyer and Wald have tried to address \cite{Iyer1}.
With this additional term, the entropy can be changed resulting in a
different entropy. We will assume that the correct action is
the one with the surface term still included and continue with the
evaluation of the original four-dimensional action.

As already emphasized, the Kodama vector
is a natural candidate to replace the Killing vector as the
generator of the relevant diffeomorphism on the horizon of a dynamic black
hole.
Using the Kodama vector to generate the diffeomorphism, the
current is
\begin{eqnarray}
J_{iab} & = & {1 \over 8 \pi }
\eta_{ab} \epsilon_{ij} ( r^2 \nabla_k \nabla^{[j}
\xi^{k]} - 2 r  \nabla_k \xi^{[j} \nabla^{k]} r  )
\\[1ex]
& = & -{1 \over 8 \pi} \eta_{ab} \nabla_i  ( r^2 \nabla_j \nabla^j r )
\end{eqnarray}
The conserved charge from the Kodama vector is then given by
\begin{equation}
Q = {1 \over 16 \pi} \eta_{ab}
\left( r^2 \nabla _j \nabla ^j r \right)
\end{equation}
Integrating this on the trapping horizon,
\begin{equation}
\oint Q = {\kappa A \over 8\pi }
\end{equation}
where $\kappa$ is the dynamical surface gravity \req{surgrav1}.
In the static case, the surface gravity is a constant on the
horizon and can be rescaled to be unity and Wald's entropy is
recovered.

The question that comes to mind is ``What is this conserved
charge?'' The normal conserved current associated with time
translations is an energy current. By replacing $\kappa$ with
\req{surgrav1}, we see that
\begin{equation}
\oint 2Q = E - 3 w V
\end{equation}
which seems to be an energy rather than entropy in general.
In the vacuum case, $w$
is zero and drops leaving just the energy.

Note also that
rescaling the surface gravity to unity is effectively assuming that
$\delta \kappa=0$, implying that
\begin{equation}
\delta \oint Q = {\kappa \delta A \over 8\pi}.
\end{equation}
The first law of the Schwarzschild black hole also takes this same
form $\delta m = \kappa \delta A / 8 \pi$.  This again suggests that
this conserved charge has more to do with an energy than an
entropy.  However, in \cite{M1}, a solution
to this problem has been suggested.

With the Wald-Iyer Noether current definition of the
entropy now adapted to a dynamical setting without the
requirement of a bifurcation surface, we can ask
questions of how the matter fields may influence this definition,
whereas before only the stationary case could be studied because of
the need for a Killing field.
However, in the above calculations, only the vacuum
and scalar field case
were studied.  In the vacuum case, the current has an extra term
that is not written down in Wald's paper, which is just the Einstein
curvature.
On shell, this is of course just zero and doesn't
affect anything.  In the scalar field case, the
term from the matter fields is the stress-energy tensor.  So
with the Einstein equations this terms again cancels.  The
question is what happens with other forms of matter?  Some initial
calculations suggest that other terms will appear in the current.

In the last two sections, we have investigated two possible
sources of entropy.  With the introductions of the trapping
horizon that is easy to define in the spherically symmetric case, we
see that in both cases of boundary terms and Noether currents
there is a possible dynamical definition of the entropy of a black hole.
This entropy is just one-quarter of the area of the trapping horizon.
This satisfies a
dynamical area law in agreement with the generalized first law
that has also been found in the spherical case \cite{Sean3}.
However, these results need to be compared with a
statistical-mechanical definition of entropy coming from a quantum
theory, which as yet has only been done in the stationary case
e.g. \cite{ABCK} or in a reduced dimensional case
e.g \cite{Carlip1}. Perhaps
the reduced dimensional action \req{reduce}
above may be simple enough to quantize and how this affects the
definition of the entropy is unclear.

\section{Acknowledgments}

SAH is supported by a European Union Science
and Technology Fellowship.
MCA is supported by NSF/JSPS Postdoctoral Fellowship for Foreign
Researchers (No.\ P97198).

\end{document}